\documentclass[twocolumn,showpacs,preprintnumbers,amsmath,amssymb]{revtex4}

\usepackage{ulem} 
\usepackage[usenames]{color}

\usepackage{graphicx}
\usepackage{dcolumn}

\newcommand{\lk}{\left( }
\newcommand{\rk}{\right)}
\newcommand{\ltk}{\left\{ }
\newcommand{\rtk}{ \right\} }
\newcommand{\ldk}{\left[ }
\newcommand{\rdk}{ \right] }

\begin{document}

\title{Skyrmions with holography and hidden local symmetry}
 


\author{Kanabu Nawa\footnote{E-mail: nawa@rcnp.osaka-u.ac.jp}, 
                 Atsushi Hosaka\footnote{E-mail: hosaka@rcnp.osaka-u.ac.jp}} 
\affiliation{Research Center for Nuclear Physics (RCNP),
             Osaka University, Mihogaoka 10-1,
             Ibaraki, Osaka 567-0047, Japan}
\author{Hideo Suganuma\footnote{E-mail: suganuma@ruby.scphys.kyoto-u.ac.jp}} 
\affiliation{Department of Physics, Graduate School of Science, 
             Kyoto University, Kitashirakawa, Sakyo, 
             Kyoto 606-8502, Japan}

\begin{abstract}
We study baryons as Skyrmions in holographic QCD
with D4/D8/$\overline{{\rm D}8}$ multi-D brane system in type IIA
superstring theory, and also in the non-linear sigma model with
hidden local symmetry (HLS).
Comparing these two models, we find that the extra-dimension
and its nontrivial curvature can largely change the role of
(axial) vector mesons for baryons in four-dimensional space-time.
In the HLS approach,
the $\rho$-meson field as a massive Yang-Mills field
has a singular configuration in Skyrmion,
which gives a strong repulsion for the baryon as a stabilizer.
When $a_1$ meson is added in this approach,
the stability of Skyrmion is lost by the cancellation
of $\rho$ and $a_1$ contributions.
On the contrary, in holographic QCD,
the $\rho$-meson field does not appear 
as a massive Yang-Mills field due to the extra-dimension
and its nontrivial curvature.
We show that the $\rho$-meson field
has a regular configuration in Skyrmion,
which gives a weak attraction for the baryon in holographic QCD.
We argue that Skyrmion with $\pi$, $\rho$ and $a_1$ mesons
become stable due to the curved extra-dimension and also the 
presence of the Skyrme term in holographic QCD.
From this result, we also discuss the features of
our truncated-resonance analysis on baryon properties
with $\pi$ and $\rho$ mesons below the cutoff scale $M_{\rm KK}\sim 1$GeV
in holographic QCD, which is compared with other 5D instanton analysis.
\end{abstract} 
\maketitle
%
\section{Introduction}

In 1961, Skyrme proposed an idea that
the baryon can be described  as a classical soliton in 
non-linear meson field theories, 
called ``Skyrmion''~\cite{Sky_mode}.
Later, in 1970's,
Skyrmion was revived in the development of large-$N_c$ QCD,
where the baryon mass $M_{\rm B}$ is found to be proportional to
the inverse of the meson-meson coupling $g_{\rm eff}$ as
$M_{\rm B}\propto 1/g_{\rm eff}$ like a ``soliton''~\cite{tH, Witten}.
Due to the Derrick theorem~\cite{Raja},
a stable soliton does not appear
only with a two-derivative term of the chiral field.
In the Skyrme model,
a four-derivative term is supplemented as a stabilizer. 
Alternately,
if one includes vector mesons like $\rho$ meson,
there appear $\rho$-$2\pi$ coupling to give a non-linear coupling of the chiral field
via $\rho$-meson propagation,
which can stabilize the soliton~\cite{Igarashi}.
In this way,
the importance of (axial) vector mesons for the Skyrmion is naturally expected.

One of 
convenient approaches for
meson phenomenologies is the ``hidden local symmetry (HLS)''~\cite{BKY},
where $\rho$ meson is introduced as the 
dynamical gauge boson of the hidden local symmetry
in the non-linear sigma model with
${\rm SU}(2)_{\rm L}\times{\rm SU}(2)_{\rm R}\times{\rm SU}(2)_{\rm V}^{\rm local}$.
Skyrmions with $\pi$, $\rho$, $a_1$, and $\omega$ mesons
are discussed in HLS and its extended models~\cite{Igarashi, KFLiu, Hosaka-Toki, E-Y}.

Recently,
Skyrmion has been revived in holographic QCD~\cite{SS}.
%
The solitonic description
for a baryon naturally appears
in holographic dual of D4/D8/$\overline{{\rm D}8}$
multi-D brane system in type IIA superstring theory~\cite{NSK, NSK1, NSK2}.
Today, 
holographic QCD is known to reproduce
many phenomenological aspects in hadron physics
{\it semi-quantitatively},
from dual classical supergravity calculations.
However,
important aspects in the
{\it difference} between flat space phenomenologies
and holographic approach have not been so much investigated.

In this paper,
we compare 
the Skyrmion in HLS and that in holographic QCD,
and find that
they have significantly different soliton solutions,
particularly in the vector-meson profiles,
due to 
the existence of the extra-dimension and its nontrivial curvature.
%
As a result, we also 
discuss the features  of 
our truncated-resonance analysis for baryons 
with $\pi$ and $\rho$ mesons below the cutoff scale $M_{\rm KK}\sim 1$GeV
in holographic QCD~\cite{NSK},
which is compared with other 5D instanton approaches.


\section{Skyrmions with hidden local symmetry \label{sec_HLS}}
%
First, we study the Skyrmion with $\rho$ meson
as the dynamical gauge boson of the hidden local symmetry (HLS) in the 
non-linear sigma model.
Non-linear sigma model with scalar manifold on coset space $G/H$
can be generally formulated by the gauge theory with linear symmetry
$G\times H_{\rm local}$ as ``gauge equivalence''~\cite{BKY}.
In case of the global chiral symmetry $G={\rm SU}(2)_{\rm L}\times{\rm SU}(2)_{\rm R}$
spontaneously broken into the isospin subgroup $H={\rm SU}(2)_{\rm V}$,
the chiral field 
$U(x)\in {\rm SU}(2)_{\rm L}\times{\rm SU}(2)_{\rm R}/
{\rm SU}(2)_{\rm V}\simeq {\rm SU}(2)_{\rm A}$
can be divided into two pieces 
with hidden variables $\xi_{\rm L}(x)$ and $\xi_{\rm R}(x)$ as 
\begin{eqnarray}
U(x)\equiv\xi_{\rm L}^{\dagger}(x)\cdot\xi_{\rm R}(x) .\label{chiral1}
\end{eqnarray}
Then one can introduce chiral symmetry transformation by 
$g_{\rm L (R)}\in {\rm SU}(2)_{\rm L(R)}$ and
hidden local symmetry transformation by
$h(x)\in {\rm SU}(2)_{\rm V}^{\rm local}$
as 
\begin{eqnarray}
\xi_{{\rm L}({\rm R})}(x)&\rightarrow& h(x)\xi_{{\rm L}({\rm R})}(x)g_{{\rm L}({\rm R})}^\dagger, \label{tra1}\\
V_\mu(x)&\rightarrow& i h(x)\partial_\mu h^{\dagger}(x)+
                                                      h(x)V_\mu(x)h^{\dagger}(x),\label{tra2}
\end{eqnarray}
where $V_\mu(x)=eV_{\mu a}(x)\frac{\tau_a}{2}$ is the ``gauge field'' of
${\rm SU}(2)_{\rm V}^{\rm local}$, 
regarded as the $\rho$ meson field in HLS.
Here $e$ is the ``gauge coupling'', 
later corresponding  to
the Skyrme parameter~\cite{Igarashi}.
By using the covariant derivatives:
\begin{eqnarray}
D_\mu\xi_{L(R)}(x)=\partial_\mu \xi_{L(R)}(x) - i V_\mu(x)\xi_{L(R)}(x), \label{cov1}
\end{eqnarray} 
several currents can be introduced as 
\begin{eqnarray}
\hat{\alpha}_{\mu \parallel}&=&(D_\mu\xi_{\rm L}\cdot\xi_{\rm L}^{\dagger}+
                                                               D_\mu\xi_{\rm R}\cdot\xi_{\rm R}^{\dagger})/2i
                                                              =\alpha_{\mu \parallel}-V_\mu, \label{current1}\\
\hat{\alpha}_{\mu \perp}&=&(D_\mu\xi_{\rm L}\cdot\xi_{\rm L}^{\dagger}-
                                                               D_\mu\xi_{\rm R}\cdot\xi_{\rm R}^{\dagger})/2i
                                                              =\alpha_{\mu \perp}, \label{current2}\\
\alpha_{\mu \parallel, \perp}
                                                      &=&(\partial_\mu\xi_{\rm L}\cdot\xi_{\rm L}^{\dagger}\pm
                                                               \partial_\mu\xi_{\rm R}\cdot\xi_{\rm R}^{\dagger})/2i, \label{current3}
\end{eqnarray}   
where $\hat{\alpha}_{\mu \parallel}$ and $\hat{\alpha}_{\mu \perp}$
are parallel and perpendicular components 
for ${\rm SU}(2)_{\rm V}^{\rm local}$, respectively.
By using these currents (\ref{current1}) and (\ref{current2}),
we can construct the chiral Lagrangian 
with HLS
in Euclidean space-time,
invariant for  
${\rm SU}(2)_{\rm L}\times{\rm SU}(2)_{\rm R}\times{\rm SU}(2)_{\rm V}^{\rm local}$
as
\begin{eqnarray}
{\cal L}_{\rm HLS}&=&{\cal L}_A+a{\cal L}_V+{\cal L}_{\rm kin},\label{L_HLS1}\\
{\cal L}_A&=&f_\pi^2 {\rm tr}(\hat{\alpha}_{\mu \perp}^2)
                            =\frac{1}{4}f_\pi^2{\rm tr}(\partial_\mu U\partial^\mu U^\dagger), \label{L_HLS_A}\\
{\cal L}_V&=&f_\pi^2 {\rm tr}(\hat{\alpha}_{\mu  \parallel}^2),\label{L_HLS_V}\\
{\cal L}_{\rm kin}&=&\frac{1}{2e^2}{\rm tr}(F_{\mu\nu}^2),\label{L_HLS_kine}
\end{eqnarray}
where $a$ is an arbitrary parameter and $f_\pi$ is the pion decay constant.
%
$\pi$ and $\rho$ mesons can interact with each other 
only through the part $a{\cal L}_{\rm V}$, where the symmetry is gauged.
Therefore $a$-parameter controls the coupling strength between $\pi$ and $\rho$. 
In this framework, 
$a=2$ is favored to
reproduce empirically successful relations
such as the KSRF relation~\cite{KSRF} and the vector-meson dominance~\cite{Sakurai}.
We have also introduced the kinetic term of $\rho$ mesons (\ref{L_HLS_kine})
with $F_{\mu\nu}=\partial_\mu V_\nu-\partial_\nu V_\mu+i[V_\mu, V_\nu]$,
as a ``dynamical pole generation'' of HLS due to the quantum effects~\cite{BKY}.

By taking the unitary gauge with $\xi_{\rm L}=\xi_{\rm R}^{-1}\equiv \xi $,
${\cal L}_{\rm V}$ can be written as
\begin{eqnarray}
{\cal L}_V=f_\pi^2{\rm tr}(J_\mu-V_\mu)^2, \label{L_HLS_V_u}
\end{eqnarray}
where
$J_\mu\equiv\alpha_{\mu \parallel}=(1/2i)(\partial_\mu U U^{\dagger}+\partial_\mu U^{\dagger}U)/
                                                                    {\rm det}(1+U)$
is the vector current of the chiral field.

For a baryon,
we now take the hedgehog Ansatz for the chiral field $U(x)$
and the $\rho$ meson field $V_\mu(x)$ as a Skyrmion: 
\begin{eqnarray}
U^{\star}(x)&=&e^{i\tau_a\hat{x}_a F(r)}  \hspace{5mm}(\hat{x}_a\equiv x_a/r, r\equiv |{\bf x}|).\label{HH_p}\\
V_0^{\star}(x)&=&0, \hspace{3mm}V_i^{\star}=e V_{ia}^{\star}(x)\frac{\tau_a}{2}=
                                     \{\varepsilon_{iab} \hat{x}_b G(r)/r\}\frac{\tau_a}{2}, \nonumber\\
                                                                                                                                   \label{HH_rho}
\end{eqnarray}
with the topological boundary condition
to ensure the baryon number $B=1$ as
\begin{eqnarray}
F(0)=\pi, \hspace{5mm}F(\infty)=0. \label{pi_bound1}
\end{eqnarray}
By substituting the Ansatz (\ref{HH_p}) and (\ref{HH_rho})
into the action,
we get the hedgehog mass of the Skyrmion~\cite{Igarashi} as follows:
\begin{eqnarray}
E[F,G]&=&4\pi\int dr \biggl[ \frac{1}{2} f_\pi^2 (r^2 F'^2 + 2\sin^2 F)\nonumber\\
                &&   + a f_\pi^2 \{G-(1-\cos F)\}^2 \nonumber\\
                &&+\frac{1}{2e^2}
                    \{2G'^2+G^2(G-2)^2/r^2 \}\biggr].\label{HHmass_HLS}
\end{eqnarray}

It is convenient to 
take the Adkins-Nappi-Witten (ANW) units for energy and length
as 
$\overline{E}\equiv \frac{1}{E_{\rm ANW}}E=\frac{2e}{f_\pi}E$
and 
$\overline{r}\equiv \frac{1}{r_{\rm ANW}}r=ef_\pi r$~\cite{ANW}.
Then (\ref{HHmass_HLS}) can be written as follows 
(below, overlines of $\overline{E}$ and $\overline{r}$ are omitted for simplicity):
\begin{eqnarray}
E[F,G]&=&4\pi\int dr \bigl[ (r^2 F'^2 + 2\sin^2 F)\nonumber\\
                && + 2a \{G-(1-\cos F)\}^2\nonumber\\
                && + \{2G'^2+G^2(G-2)^2/r^2 \}\bigr],\label{HHmass_HLSanw}
\end{eqnarray}
where the parameter dependence appears only in $E_{\rm ANW}$, 
$r_{\rm ANW}$, and coupling strength $a$.

\begin{figure}[t]
\begin{center}
       \resizebox{75mm}{!}{\includegraphics{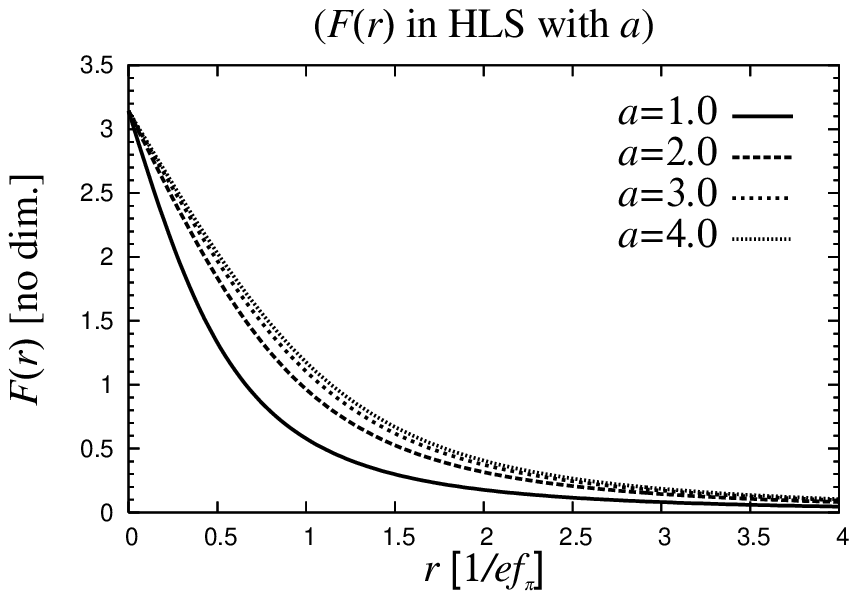}}\\ 
           \vspace{6mm}
       \resizebox{75mm}{!}{\includegraphics{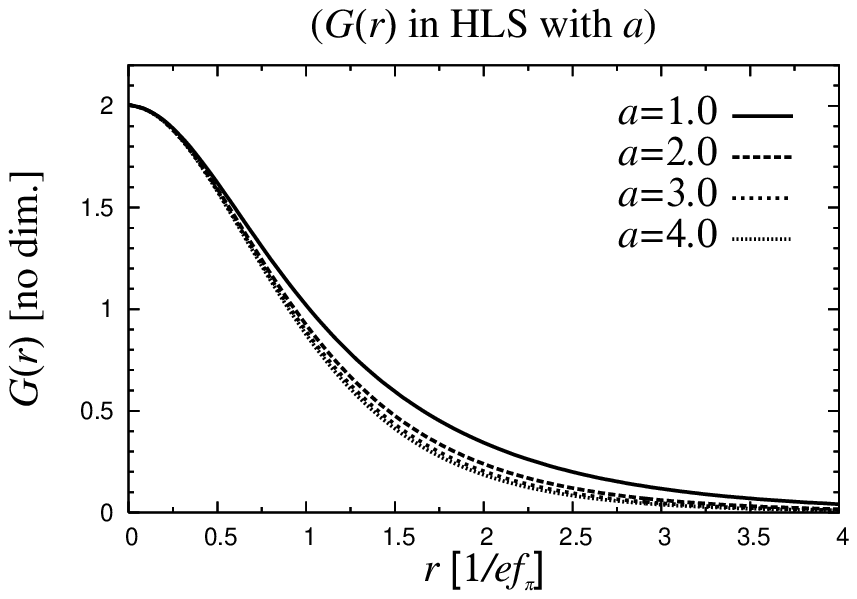}}
\end{center}
\caption{$F(r)$ and $G(r)$ of Skyrmion in hidden local symmetry (HLS) approach
with various values of $a$-parameter in (\ref{L_HLS1}).}
\label{HLS_num1}
\end{figure}
\begin{figure}[t]
  \begin{center}
       \resizebox{75mm}{!}{\includegraphics{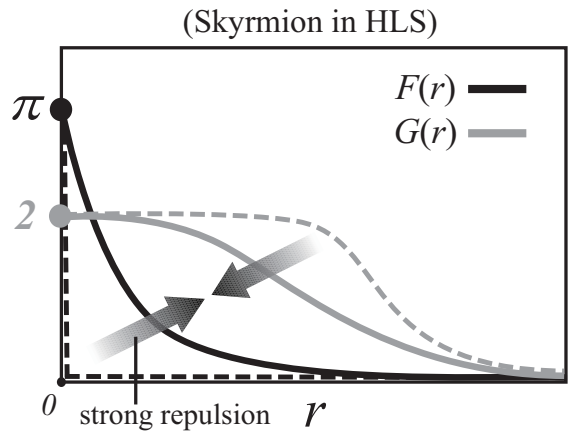}}\\
  \end{center}
\caption{Schematic figure of Skyrmion in HLS.
Dashed black and grey lines show the configurations 
of $F(r)$ and $G(r)$ without interactions, respectively.
$G(r)$ appears as broad singular configuration with the boundary $G(0)=2$.}
  \label{HLS_sche1}
\end{figure}

By solving the Eular-Lagrange  equations for $F(r)$ and $G(r)$ 
in the minimization of the energy (\ref{HHmass_HLSanw}),
we find a stable Skyrmion as shown in Fig.~\ref{HLS_num1}. 
One should note that $G(r)$ has a non-zero value at the origin:
$G(0)=2$.
Therefore, 
to keep the total energy finite,
$\rho$ meson field (\ref{HH_rho}) appears as
a singular configuration with $V_i\propto \frac{1}{r}$ around $r=0$,
having a spatially ``broad'' extension 
to gain the kinetic-energy in (\ref{HHmass_HLSanw}).
Then such broad $\rho$ meson field
will pull up the pion field
along the radial direction, which is the origin of the
{\it strong repulsion} for the Skyrmion with a finite size (see a schematic plot in Fig.~\ref{HLS_sche1}).
In fact, by increasing the coupling strength $a$,
pion field $F(r)$ is pulled up to outside of baryon, while
$\rho$ meson field $G(r)$ is pulled down to inside.
%
Furthermore,
without the $\rho$ meson field,
the leading term of the chiral field ${\cal L}_A$
alone does not support the stable Skyrmion due to the
Derrick theorem~\cite{Raja}.
Therefore, $\rho$ meson field as the broad singular configuration
in the massive Yang-Mills sector of HLS
plays the essential role as the stabilizer of the Skyrmion.
(Such singular solution  was first analyzed in the Yang-Mills theory in
Ref.~\cite{WY}.)

Now, dividing the chiral field $U(x)$ in (\ref{chiral1})
into three hidden pieces as $U(x)=\xi_1\xi_2\xi_3$,
one can introduce $\rho$ and $a_1$ mesons
for two independent hidden symmetries, 
which we now call the ``HLS$_2$ model''.
However,
the repulsive effect of $\rho$ meson was found to be canceled by $a_1$ meson in HLS$_2$,
giving the catastrophic instability of Skyrmions.
Then, the $\omega$ meson was additionally introduced
as the origin of the strong repulsion for the baryons~\cite{KFLiu, Hosaka-Toki}.

\section{Skyrmions with holography \label{sec_HOL}}
Next, we study the Skyrmion in holographic QCD.
Let us begin with the non-abelian Dirac-Born-Infeld (DBI) action of
D8 branes with D4 supergravity background,
to be dual of strong-coupling large-$N_c$ QCD.
After dimensional reductions,
one gets the five-dimensional Yang-Mills action~\cite{SS} as 
\begin{eqnarray}
S_{\rm D8}^{\rm DBI}&=&\kappa\int d^4x dz {\rm tr}
\ltk \frac{1}{2}K(z)^{-1/3}F_{\mu\nu}^2+K(z)F_{\mu z}^2\rtk\nonumber\\ 
                                                   &&+O(F^4), \label{5YM}
\end{eqnarray}
with $\kappa=\lambda N_c/216\pi^3$, and $\lambda$ is the 'tHooft coupling.
$K(z)=1+z^2$ 
expresses the nontrivial curvature in the fifth dimension $z$.
After proper mode expansions for the five-dimensional gauge field
$A_M(x,z)$ ($M=0\sim 3, z$) with a complete orthogonal basis in the
fifth dimension $z$,
one can get 
an action written by physical meson degrees of freedom 
with definite parity and G-parity
in four-dimensional space-time~\cite{NSK}.
%
The resulting Euclidean effective
action 
for chiral field $U(x)$, $\rho$ meson field $V_\mu(x)$ and 
infinite number of (axial) vector mesons is as follows:
\begin{eqnarray}
S_{\rm D8}^{\rm DBI}\hspace{-1.6mm}&=&\hspace{-1.6mm}
\frac{f_\pi^2}{4}\int d^4x {\rm tr} (L_\mu^2)
-\frac{1}{32e^2}\int d^4x {\rm tr}[L_\mu, L_\nu]^2\nonumber\\
&&
+\frac{m_\rho^2}{e^2}\int d^4 x {\rm tr}(V_\mu^2)
+\frac{1}{2e^2}\int d^4 x {\rm tr}(\partial_\mu V_\nu-\partial_\nu V_\mu)^2\nonumber\\
&&
+i\frac{1}{2e^2}2g_{3\rho}\int d^4 x {\rm tr}\{ (\partial_\mu V_\nu-\partial_\nu V_\mu)[V_\mu, V_\nu]\}\nonumber\\
&&
-\frac{1}{2e^2}g_{4\rho}\int d^4 x {\rm tr}[V_\mu, V_\nu]^2
+S_{\pi\mbox{-}\rho}+S_{a_1, \rho', a'_1, \rho'', \cdots}, \nonumber\\
&&
\label{action_HOL1}
\end{eqnarray} 
where $L_\mu=(1/i)U^\dagger \partial_\mu U$ is one-form of the chiral field,
$S_{\pi\mbox{-}\rho}$ interaction terms between $\pi$ and $\rho$, and $S_{a_1, \rho', a'_1, \rho'', \cdots}$
contributions from higher excited resonance of (axial) vector mesons
(see Ref.~\cite{NSK} for details;
$\frac{1}{e}V_\mu$, $eg_{3\rho}$, and $e^2g_{4\rho}$ in (\ref{action_HOL1})
correspond to $V_\mu$, $g_{3\rho}$, and $g_{4\rho}$ in Ref.~\cite{NSK}
as just field redefinitions).
There appear infinite number of coupling constants like
pion decay constant  $f_\pi$,
$\rho$-meson mass $m_\rho$,
Skyrme parameter $e$,
three-point coupling $g_{3\rho}$,
four-point coupling $g_{4\rho}$,
and other coupling constants in $S_{\pi\mbox{-}\rho}$ and $S_{a_1, \rho', a'_1, \rho'', \cdots}$.
%
However,
holographic QCD has just two independent parameters: $\kappa\propto N_c$ and $M_{\rm KK}$
as the Kaluza-Klein compactification scale of D4 branes.
Therefore, by taking two experimental inputs like
$f_\pi=92.4$MeV and $m_\rho=776.0$MeV,
all the coupling constants are uniquely determined.

\begin{figure}[ht]
  \begin{center}
       \resizebox{72mm}{!}{\includegraphics{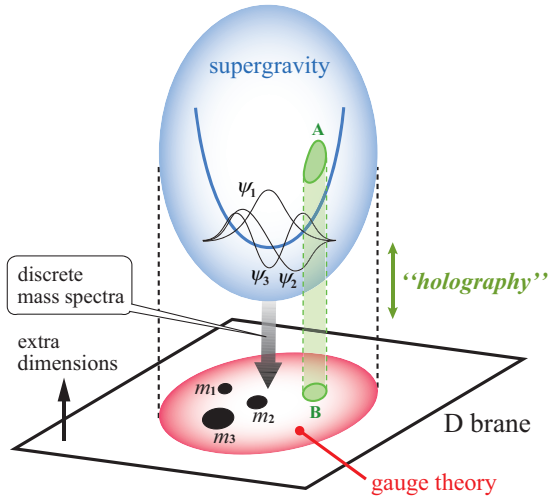}}\\
  \end{center}
\caption{Schematic figure of holographic duality
between gauge theory as QCD on D brane and supergravity around D brane.
Masses of mesons $m_{n=1,2,3,\cdots}$
are given by the oscillation of meson wave functions $\psi_n(z)$
in the extra-dimension $z$~\cite{NSK}.
Therefore,
the curvature in the extra-dimension around D brane
is essential to give the {\it discrete} meson mass spectra as QCD on D brane.
Then, regions A and B have different curvatures, and
the gauge symmetry on A will be distorted on B except for the gauge symmetry of QCD,
giving the relation $g_{3\rho}^2/g_{4\rho}\neq 1$.
}
  \label{dual_sche1}
\end{figure}

Here we remark several interesting
features of the meson effective action (\ref{action_HOL1}).
First, 
the Skyrme term $\int d^4 x {\rm tr}[L_\mu, L_\nu]^2$
naturally appears as a stabilizer of the Skyrmion.
In this sense, the Skyrme soliton picture is supported by the holographic approach~\cite{NSK}.
Second,
the relation $g_{3\rho}^2/g_{4\rho}=1$ does {\it not} hold.
In fact, 
due to the existence of the fifth dimension $z$ and
its nontrivial curvature,
the ratio $g_{3\rho}^2/g_{4\rho}$ deviates from unity
in holographic QCD as 
\begin{eqnarray}
\frac{g_{3\rho}^2}{g_{4\rho}}=
\frac{\ldk \kappa\int_{-\infty}^{+\infty} dz K(z)^{-1/3}\psi_1(z)^3\rdk^2}
{\kappa\int_{-\infty}^{+\infty}  dz K(z)^{-1/3}\psi_1(z)^4}
\simeq 0.90, \label{ratio_HOLO}
\end{eqnarray}
where $\psi_1(z)$ is 
a proper basis corresponding to
the $\rho$-meson wave-function 
in $z$-dimension~\cite{NSK}.
%
Note that, with
$\psi_1(z)\propto \kappa^{-1/2}$, 
the value $g_{3\rho}^2/g_{4\rho}$ in (\ref{ratio_HOLO})
is independent of the parameters, $\kappa$ and $M_{\rm KK}$, 
in holographic QCD.
This implies that the kinetic term of the $\rho$ meson
differs from Yang-Mills-type, i.e.,
the $\rho$-meson field does {\it not} appear as
a massive Yang-Mills field in four-dimensions
due to the curved $z$-dimension,
which should be compared with the case of HLS with $g_{3\rho}^2/g_{4\rho}=1$ 
in Eq.~(\ref{L_HLS_kine}). 
In fact,
the masses $m_n$ of (axial) vector mesons 
(e.g., $m_1=m_\rho$, $m_2=m_{a_1}$, $m_3=m_{\rho '}$, $m_4=m_{a'_1}$, etc.)
are given by the eigenvalues $\lambda_n$ of the meson wave function $\psi_n(z)$ 
in $z$-dimension as $m_n^2=\lambda_n$~\cite{NSK}.
In this sense, the existence of the curvature $K(z)$ in $z$-dimension
is essential to give the {\it discrete} mass spectra for mesons
with $m_\rho<m_{a_1}<m_{\rho '}<m_{a'_1}<\cdots$,
similarly to the Kaluza-Klein mechanism with the dimensional reductions
(see Fig.~\ref{dual_sche1}).
(If extra-dimensions have no curvatures,
all the modes belong to the continuum as plane waves,
which contradicts meson properties.)
In the following, we propose that any deviation of the ratio $g_{3\rho}^2/g_{4\rho}$
from unity 
can drastically change the hadron properties in four-dimensional space-time.

Now we take only $\pi$ and $\rho$ mesons below $M_{\rm KK}\sim 1$GeV
as the ultraviolet cutoff scale in holographic approach.
This strategy is called the ``truncated-resonance model'' for the baryons~\cite{NSK}.
By substituting the hedgehog configuration Ansatz (\ref{HH_p}) and (\ref{HH_rho}) 
in the action (\ref{action_HOL1}),
we get the hedgehog mass of the Skyrmion in ANW unit as 
\begin{eqnarray}
E[F,G]\hspace{-1.6mm}&=&\hspace{-1.6mm}4\pi \int dr r^2 \varepsilon[F,G]\label{mass1}\\
r^2 \varepsilon[F,G]\hspace{-1.6mm}&=&\hspace{-1.6mm}
(r^2 F'^2+2\sin^2F)
+\sin^2 F\lk 2F'^2+\sin^2 F/r^2 \rk\nonumber\\
&&
+2\frac{m_\rho^2}{e^2 f_\pi^2} G^2
+(4G^2/r^2+2G'^2) \nonumber\\
&&
-2g_{3\rho}(2G^3/r^2)
+g_{4\rho}(G^4/r^2)
+r^2\varepsilon_{\pi\mbox{-}\rho}[F,G],\nonumber\\
&&
\label{energy_density1}
\end{eqnarray}
where $\varepsilon_{\pi\mbox{-}\rho}$ is the contribution from 
$S_{\pi\mbox{-}\rho}$ in (\ref{action_HOL1}) (see Ref.~\cite{NSK} for details;
$\frac{1}{2e}G$ in (\ref{energy_density1}) corresponds to $G$ in Ref.~\cite{NSK}
as just field redefinition
with hedgehog Ansatz (\ref{HH_rho})).

\begin{figure}[t]
\begin{center}
       \resizebox{75mm}{!}{\includegraphics{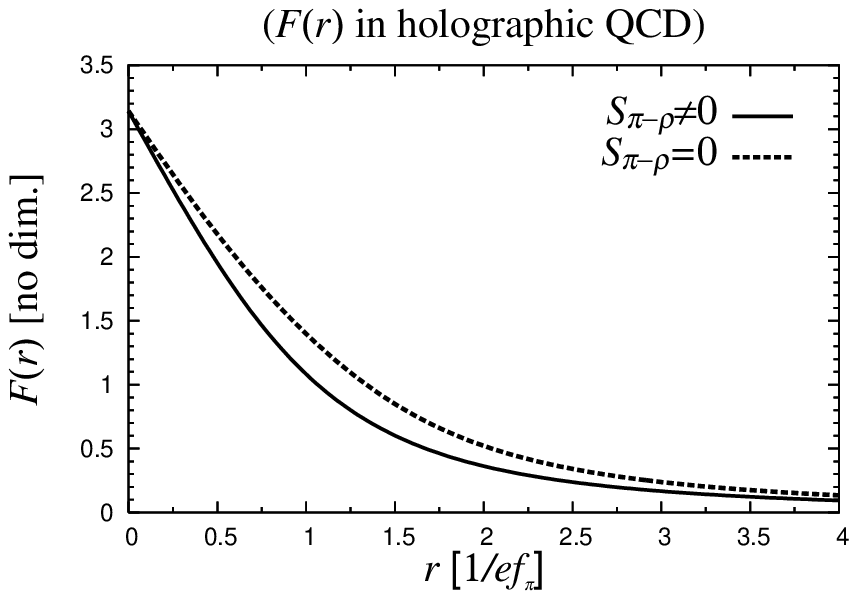}} \\
           \vspace{6mm}
       \resizebox{75mm}{!}{\includegraphics{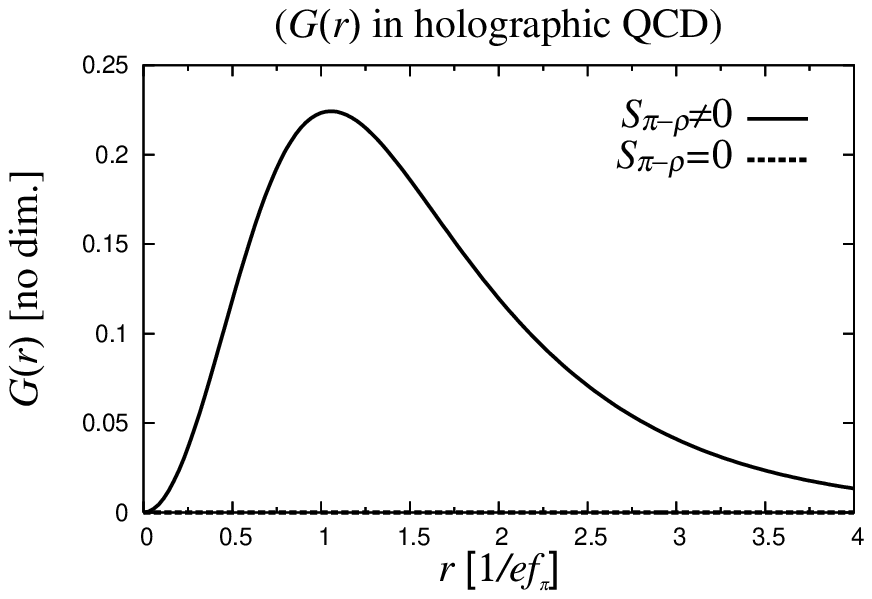}}
\end{center}
\caption{$F(r)$ and $G(r)$ of Skyrmion in holographic QCD.
Dashed lines with label ``$S_{\pi\mbox{-}\rho}=0$'' show the configurations without interactions
between $\pi$ and $\rho$ in (\ref{action_HOL1}).}
\label{HOL_num1}
\end{figure}
\begin{figure}[t]
  \begin{center}
       \resizebox{75mm}{!}{\includegraphics{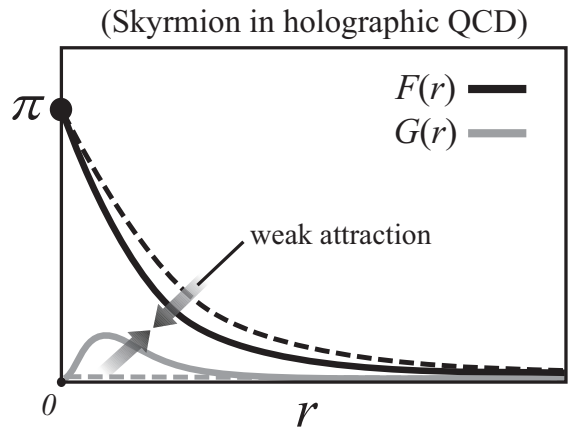}}\\
  \end{center}
\caption{Schematic figure of Skyrmion in holographic QCD.
Dashed black and grey lines show the configurations 
of $F(r)$ and $G(r)$ without interactions, respectively.
$G(r)$ appears as regular configuration in the core region of a baryon
with the boundary $G(0)=0$.}
  \label{HOL_sche1}
\end{figure}

By solving the Eular-Lagrange equations for $F(r)$ and $G(r)$,
we find a stable Skyrmion as shown in Fig.~\ref{HOL_num1}.
One should note that $\rho$ meson field appears as a regular configuration
with $G(0)=0$, not as the broad singular one with $G(0)=2$ in Sec.~\ref{sec_HLS}.
Therefore, $\rho$-meson field appearing in the core region of the baryon
will pull down the pion field to inside of the baryon,
giving the {\it weak attraction} for the baryon with a slight shrinkage of its total size~\cite{NSK}
(see a schematic plot in Fig.~\ref{HOL_sche1}).
This result is clearly different from the case of HLS in Sec.~\ref{sec_HLS},
where $\rho$ meson field gives the strong repulsion for the baryon.
We find that such a remarkable difference comes 
from the value $g_{3\rho}^2/g_{4\rho}$. 
In fact,  substituting $G(r)=\alpha$ (constant) near the origin $r=0$,
the energy density in (\ref{energy_density1}) is dominated
by $(\partial_\mu V_\nu-\partial_\nu V_\mu)^2$, 
three-point and four-point $\rho$-meson
coupling terms as
\begin{eqnarray}
(4\alpha^2 r^{-2})-2g_{3\rho}(2\alpha^3 r^{-2})+
g_{4\rho} (\alpha^4r^{-2}), \label{dominate_1}
\end{eqnarray}   
%
which is ultraviolet-divergent in the integration over $r$.
Therefore, for $G(r)=\alpha$ to be realized near $r=0$ as a finite energy configuration,
the coefficient of $r^{-2}$ in (\ref{dominate_1}) should vanish as
\begin{eqnarray}
\alpha^2(g_{4\rho}\alpha^2-4g_{3\rho}\alpha+4)=0. \label{cond1}
\end{eqnarray}
Furthermore,
the Eular-Lagrange equation of $G(r)$ can also be dominated near $r=0$ 
by the terms (\ref{dominate_1})
as
\begin{eqnarray}
\frac{1}{4\pi}\ldk    \frac{\delta E[F,G]}{\delta G(r)}
-\frac{d}{dr}\ltk \frac{\delta E[F,G]} {\delta G'(r)}\rtk\rdk_{r\rightarrow 0}\nonumber\\
=4\alpha (g_{4\rho}\alpha^2-3g_{3\rho}\alpha+2)r^{-1}=0. \label{cond2}
\end{eqnarray}
Therefore, 
for $G(r)$ to 
take a non-zero value
near $r=0$ as a finite energy soliton solution,
the following must be satisfied as the necessary condition 
from (\ref{cond1}) and (\ref{cond2}) as
\begin{eqnarray}
\alpha\neq 0\hspace{5mm}  {\rm and}  \hspace{5mm}\alpha=2/g_{3\rho} 
                            \hspace{5mm}  {\rm and}  \hspace{5mm}  g_{3\rho}^2/g_{4\rho}=1, \label{cond3}
\end{eqnarray}
otherwise, $\alpha=0$.
Actually, 
(\ref{cond3}) is the condition of the Yang-Mills field,
reproducing the results of HLS in Sec.~\ref{sec_HLS}
with $g_{3\rho}=1$ and $G(0)=2$.
In case of holographic QCD, $g_{3\rho}^2/g_{4\rho}\neq 1$, and therefore
only the regular $\rho$ meson configuration with $G(0)=0$ is allowed.
As a whole, even any deviation of the ratio
$g_{3\rho}^2/g_{4\rho}$ from unity
due to
the curved extra-dimension 
can drastically change 
the baryon profiles:
for $g_{3\rho}^2/g_{4\rho}=1$ the $\rho$ meson with a strong repulsion, and
for $g_{3\rho}^2/g_{4\rho}\neq 1$ with a weak attraction for the baryon.


Here we note several expected features.
Even if $a_1$ meson is included for the Skyrmion analysis in holographic QCD,
there is {\it no} catastrophic instability as in the case of HLS$_2$ in Sec.~\ref{sec_HLS} 
by the following reasons:
First,
the Skyrme term naturally appears from the DBI action of D8 brane as a stabilizer of the Skyrmion.
Second,
because of the relation $g_{3\rho}^2/g_{4\rho}\neq 1$ 
in holographic QCD,
$\rho$ meson has weak attraction, not strong repulsion for the baryon.
Therefore, even if the $\rho$-meson contributions could be canceled by $a_1$ meson,
there is no catastrophic instability of Skyrmion into zero size
in the presense of the Skyrme term.

Recently,
the baryon is also discussed as the holonomy of 
an {\it instanton} in the five-dimensional gauge theory on
D8 branes with D4 supergravity background~\cite{HRYY,HSSY,d_Inst,PW}.
Since the instanton is introduced before the mode expansions of five-dimensional gauge 
field $A_M(x,z)$,
one would expect that
it is composed by pion and infinite tower of 
(axial) vector mesons $\rho$, $a_1$, $\rho'$, $a'_1$, $\rho''$$\cdots$.
Actually, the DBI action of D8 brane is known to lead the instability of instanton
into zero size, so that the inclusion of the Chern-Simons (CS) action is claimed as the stabilizer~\cite{HRYY,HSSY,d_Inst,PW},
though the CS sector is, in the 'tHooft coupling expansion,
higher order contribution 
relative to the DBI sector.
Then, one might relate the instability of instanton
with the instability of Skyrmion with $\pi$, $\rho$ and $a_1$ mesons 
as in the case of HLS$_2$ in Sec.~\ref{sec_HLS}.
However, we can argue that
Skyrmion with $\pi$, $\rho$ and $a_1$ mesons have {\it no} catastrophic instability
in holographic QCD due to the curved extra-dimension.
%
Once the stability is established,
additional inclusion of the CS sector, corresponding to 
the effects of the $\omega$ meson~\cite{MKWW},
does  not affect the present discussions of the stability 
with the $a_1$ meson in holographic QCD.

Here, it is noted that holographic QCD has the ultraviolet cutoff scale
$M_{\rm KK}\sim 1$GeV. In fact, there exist infinite number of non-QCD Kaluza-Klein
modes in the $S^1$-compactification of D4 branes. 
Therefore, 
the D4/D8/$\overline{{\rm D}8}$
multi-D brane system can be regarded as QCD below the $M_{\rm KK}$ scale. Indeed, the
appearance of $M_{\rm KK}$ is essential to reproduce the non-SUSY nature in QCD,
which breaks the conformal invariance to give color confinement and 
chiral-symmetry breaking. 
Now, 
in the formation of the instanton in the
five-dimensional space-time, the infinite number of non-QCD modes 
above $M_{\rm KK}$ are also
included. 
Therefore, 
one should be careful to their possible artifacts, 
which may
cause the instability of the instanton. 
In our study, we truncate the meson
resonances at the $M_{\rm KK}$ scale to include only the relevant QCD modes at the
level of the classical supergravity.

If one would like to see a fine structure of the baryon and also its 
excitations, a larger $M_{\rm KK}$ is to be taken. 
In such a case, heavier meson resonances are also 
systematically treated in a way consistent with the cutoff scale. 
In fact, the upper bound of $M_{\rm KK}$ comes from 
the local approximation of the fundamental strings on the D4 branes, 
which is compactified with a radius
$M^{-1}_{\rm KK}$. 
Therefore, if one can, e.g., include the effect of
finite string length in the action (\ref{5YM}), $M_{\rm KK}$ might be taken sufficiently
large with fixed $\Lambda_{\rm QCD}$~\cite{NSK2,pS}. 
In this sense, the instability 
of the instanton into zero size might imply a need of more consistent treatment with
quantum corrections and string length at short distance.

Such a situation may resemble that of the strong-coupling expansion in QCD
with
the plaquette lattice gauge action~\cite{[1]}. 
For a strong-coupling with large $g^2$, 
the expansion in powers of $1/g^2$ gives a basic picture of quark
confinement~\cite{[2]} and chiral-symmetry breaking~\cite{[3]}, 
providing also an analytical method to calculate hadron properties~\cite{[3]}. 
The strong-coupling QCD, however, corresponds to a large lattice spacing $a$,
and has a limitation on its spatial resolution. 
Then, if one includes
higher-order terms of $1/g^2$ in the cluster expansion~\cite{[1]}, a finer
structure of the theory becomes gradually visible.

Now, for a larger $M_{\rm KK}$, 
the relation $g^2_{3\rho}/g_{4\rho} \neq 1$ 
still holds due to the curvature in the extra-dimensions, and the Skyrme term 
generally exists as a stabilizer of the Skyrmion in dual of QCD,
which is manifest in the classical limit
as in the action (\ref{5YM}). 
With these considerations,
we can expect 
the stability of the Skyrmion
in the present study even for the larger $M_{\rm KK}$, 
which will be more rigorously discussed elsewhere 
in near future.

\section{Summary and outlook\label{SO}}
We have studied baryons as the Skyrmions 
in hidden local symmetry (HLS) and also in holographic QCD.
We conclude that the relation $g_{3\rho}^2/g_{4\rho}\neq 1$
due to the extra-dimension and its nontrivial curvature
can drastically change the roles of (axial) vector mesons
for the baryons in four-dimensions.
In fact,
the effective action density of Yang-Mills shape as
$F_{MN}F^{MN}$ ($M,N=0\sim 3,z$)
on the probe D8 brane can be inevitably distorted
through the projection into the flat four-dimensional space-time,
giving the relation $g_{3\rho}^2/g_{4\rho}\neq 1$.
Then, we also 
discuss the features of our truncated-resonance analysis
for baryons
with $\pi$ and $\rho$ mesons
below $M_{\rm KK}\sim 1$GeV 
in holographic QCD.

Our results are not limited in the case of holographic model with D4/D8/$\overline{{\rm D}8}$
multi-D brane configurations as the Sakai-Sugimoto model.
Even if one starts from any muti-D brane configurations with gauge theory
on their surface, 
and also if one rely on the Gauge/Gravity duality,
there is no reason for the (axial) vector mesons to appear 
as the Yang-Mills field
with the constraint $g_{3\rho}^2/g_{4\rho}= 1$,
because of 
the curved extra-dimension.
%
Note here that, in the holographic framework,
the curvature in the extra-dimension
is needed to give the discrete mass spectra of hadrons
in the projected four-dimensional theory as QCD.
Furthermore, also from the phenomenological point of view in four dimensions,
the constants $g_{3\rho}^2$ and $g_{4\rho}$ can be easily shifted 
at the quantum level, 
spoiling the condition $g_{3\rho}^2/g_{4\rho}= 1$.
As a whole, our results in Sec.~\ref{sec_HOL} should be more general relative to the case of HLS in Sec.~\ref{sec_HLS}.

Of course, HLS has an interesting physical meaning.
Actually, dividing the chiral field (\ref{chiral1}) into 
a sequence of products as
$U(x)=\xi_1\xi_2\cdots\xi_{N+1}$,
one can introduce gauge field $A_\mu^k (x)$$(k=1,2,\cdots,N)$ for each gauge symmetry acting at each cutting point.
If one takes the limit $N\rightarrow \infty$ to include infinite tower of (axial) vector mesons,
one can naturally introduce the five-dimensional gauge field $A_\mu(x,z)$ by shifting the discrete index $k$
into continuous variable $z$.
This strategy is called the dimensional deconstruction model of QCD, 
or bottom-up holographic approach~\cite{Son}.
In this sense, a fact that certain local symmetry is {\it hidden}
within the chiral field~\cite{BKY}
has implied the existence of larger gauge symmetry extending 
to the {\it extra-dimensions},
which corresponds to the theoretical background of HLS today.

However, HLS does not have
the concept of curvature or gravity in the extra-dimension.
In this paper, we suggest that such a curved extra-dimension can
drastically change the view of our daily life in four-dimensional space-time.

\section*{Acknowledgements}
K.N. thanks 
Prof. Masayasu Harada for his fruitful communications.
A.H. and H.S. are supported in part by the Grant for Scientific Research
[(C) No.19540297, (C) No.19540287] from the Ministry of Education,
Culture, Science and Technology (MEXT) of Japan.
This work is supported by the Global COE Program,
``The Next Generation of Physics, Spun from Universality and Emergence''
at Kyoto University.

\end{document}